\documentclass[aps,onecolumn,groupedaddress,showpacs,notitlepage,nofootinbib]{revtex4-1}

\usepackage{graphicx}
\usepackage{dcolumn}
\usepackage{bm}
\usepackage{amsmath}
\usepackage{amsthm}
\usepackage{multirow}
\usepackage{enumerate}
\usepackage{slashed}

\usepackage{amsfonts}
\usepackage{ifthen}
\usepackage{psfrag}
\usepackage{slashed}
\usepackage{hyperref}
\usepackage{gensymb}
\usepackage[utf8]{inputenc}
\usepackage{color}
\usepackage{subfigure}

\allowdisplaybreaks[4]
\newcommand{\GeV}{\text{GeV}}

\newcommand{\be}{\begin{equation}}
\newcommand{\ee}{\end{equation}}

\newcommand{\met}{\slashed{E}_T~}

\begin{document} 

\title{Z-pole test of effective dark matter diboson interactions at the CEPC}

\author{Mingjie Jin$^{1,2}$}
\email{jinmj@ihep.ac.cn}
\author{Yu Gao$^1$}
\email{gaoyu@ihep.ac.cn}

\affiliation{$^1$ Key Laboratory of Particle Astrophysics, Institute of High Energy Physics, Chinese Academy of Sciences, Beijing, 100049, China}
\affiliation{$^2$ School of Physical Sciences, University of Chinese Academy of Sciences, Beijing, 100049, China}

\begin{abstract}

In this paper we investigate the projected sensitivity to effective dark matter (DM) - diboson interaction during the high luminosity $Z$-pole and 240 GeV runs at the proposed Circular Electron Positron Collider (CEPC). The proposed runs at the 91.2 GeV $e^+e^-$ center of mass energy offers an interesting opportunity to probe effective dark matter couplings to the $Z$ boson, which can be less stringently tested in non-collider searches.
We investigate the prospective sensitivity for dimension 6 and dimension 7 effective diboson operators to scalar and fermion dark matter. These diboson operators can generate semi-visible $Z$ boson decay, and high missing transverse momentum mono-photon signals that can be test efficiently at the CEPC, with a small and controllable Standard Model $\gamma\bar{\nu}\nu$ background. A projected sensitivity for effective $\gamma Z$ coupling efficient $\kappa_{\gamma Z}< (1030$ GeV$)^{-3}$, $(1970$ GeV$)^{-3}$ for scalar DM, $\kappa_{\gamma Z}< (360$ GeV$)^{-3}$, $(540$ GeV$)^{-3}$ for fermion DM are obtain for 25 fb$^{-1}$ and 2.5 ab$^{-1}$ $Z$-pole luminosities assuming the optimal low dark matter mass range. In comparison the effective DM-diphoton coupling sensitivity $\kappa_{\gamma \gamma}< (590$ GeV$)^{-3}$ for scalar DM, $\kappa_{\gamma \gamma}< (360$ GeV$)^{-3}$ for fermion DM are also obtained for a 5 ab$^{-1}$ 240 GeV Higgs run. We also compare the CEPC sensitivities to current direct and indirect search limits on these effective DM-diboson operators.
\end{abstract}

\maketitle


\section{Introduction}
\label{sect:intro}

Astrophysical~\cite{Zwicky:1937zza, Clowe:2003tk} and cosmological~\cite{Spergel:2006hy}\cite{Ade:2015xua} evidences indicate the existence of dark matter(DM) as a major component of our Universe. From a particle physics point of view, a DM candidate particle can emerge from various theories beyond the Standard Model (SM). A weakly interacting DM particle at the electroweak mass scale (WIMP) is the most popular candidate for its natural prediction of today's thermal relic matter density. Yet a WIMP does not necessarily interact with weak forces themselves or directly couple to SM particles. Instead, its interaction to the SM can be mediated by other new physics particles that participate in SM interactions. These SM-interacting mediators can be efficiently searched for at colliders~\cite{bib:CMS_exortica, bib:ATLAS_exortica} and are often constrained to be massive. For heavy mediators, one can adopt the effective theory and let the dark matter obtain effective couplings to the SM particles. For instance, mediators that carry SM gauge interaction charges may induce effective DM coupling to SM gauge bosons at loop level. A model independent approach is to study general forms of high-order effective DM-SM operators and their phenomenology in indirect, direct detection and collider searches.

At colliders, effective interactions allow DM particles to be produced as missing transverse energy (MET) by in association with a mono-jet~\cite{bib:monojet} or single gauge boson~\cite{bib:monoV, Crivellin:2015wva} final states. The effective DM coupling gauge bosons has been actively searched for at the LHC and the strongest limits comes from the mono-photon channel~\cite{bib:LHC_monophoton}. Direct detection also give significant constraint on the DM's effective coupling to gauge bosons, and particularly in effective photonic couplings due the low momentum transfer in the nucleus - DM collision process. To the lowest dimension, effective DM - gauge boson operators couple the DM bilinears to one SM gauge boson field. Due to their lower dimensional dependence on the interaction scale, the constraints on single gauge boson - DM operators become increasingly stringent. We consider the higher dimensional diboson operators that are currently less constrained in this study. 

The proposed high luminosity runs at the future Circular Electron Position Collider (CEPC) offer a unique opportunity to DM effective couplings to the $Z$ boson. The $Z$-pole runs in particular, with projected 10$^9$ (giga-$Z$) or 10$^{11}$ (tera-$Z$) integrated on-shell $Z$ luminosities, will greatly improve the test for an effective $Z$ coupling to dark matter. Effective DM-diboson coupling to $\gamma Z$, $Z Z$ leads to resonance-enhanced production of DM and an associated photon for a DM lighter than one half of $Z$ mass. While a mono-photon final state does not reconstruct back to the $Z$ mass, the single photon with large transverse momentum and recoiling MET offer a clean test against a relatively small SM $\nu\nu\gamma$ background.

In this work, we consider the effective theory DM-diboson interaction to $\gamma$ and $Z$ and study the sensitivity at the CEPC's $Z$-pole and 240 GeV runs. We briefly discuss the effective operators and the induced photon spectra in Section~\ref{sect:operators} and~\ref{sect:z_spec}. We analyze the CEPC mono-photon signals in Section~\ref{sect:cepc}. Comparisons between CEPC, direct and indirection searches are given in Section~\ref{sect:limits} and then we conclude in Section~\ref{sect:conclusion}.

\section{Effective diboson operators}
\label{sect:operators}

Standard Model gauge singlet DM can obtain loop-level coupling to the SM gauge bosons if they couple to heavy new physics state that are charged under SM gauge interactions. A comprehensive list of high-dimensional operators are discussed in Ref.~\cite{Rajaraman:2012fu, Chen:2013gya, Liu:2017zdh}. To the lowest order, such operators would also let the DM couple to one gauge boson, for instance, the electromagnetic dipole~\cite{Barger:2010gv, Geytenbeek:2016nfg} and anapole~\cite{Ho:2012bg} interactions. DM - single boson interactions lead to significant direct-detection signals and hence are stringently constrained~\cite{Banks:2010eh, DelNobile:2014eta}. Here we consider the higher order DM-diboson operators of dimension 6 and 7. With a focus on the production from the $e^+e^-$ collision, we only consider the coupling to electroweak gauge fields $W,B$, 
\begin{eqnarray}
&\mathcal{L}_1&=\kappa_1\phi^{\ast}\phi B^{\mu\nu}B_{\mu\nu}+\kappa_2\phi^{\ast}\phi W^{a,\mu\nu}W^a_{\mu\nu}~~~~~~~~~~(D=6),\label{eq:l0}\\
&\mathcal{L}_2&=\kappa_1\bar{\chi}\chi B^{\mu\nu}B_{\mu\nu}+\kappa_2\bar{\chi}\chi W^{a,\mu\nu}W^a_{\mu\nu}~~~~~~~~~~~~~(D=7),\label{eq:l1}\\
&\mathcal{L}_3&=\kappa_1\bar{\chi}i\gamma_5\chi B^{\mu\nu}\tilde{B}_{\mu\nu}+\kappa_2\bar{\chi}i\gamma_5\chi W^{a,\mu\nu}\tilde{W}^a_{\mu\nu}~~~~~(D=7),\label{eq:l2}
\end{eqnarray}
where we denote the spin-0 and spin-$1/2$ DM fields as $\phi$ and $\chi$, which are singlets under SM interactions. $B_{\mu\nu}$ and $W_{\mu \nu}$ are the SM $U(1)_Y$, $SU(2)_L$ gauge field strengths (FS). The CP-odd field strength dual $\tilde{B}$ and $\tilde{W}$ would couple to the pseudo-scalar product of the dark matter $\bar{\chi}i\gamma^5\chi$. $\kappa$ is the effective coupling coefficient for each term and it is of dimension -2 or -3. After electroweak symmetry breaking the operators can be written for the physical $\gamma,Z$ and $W$ fields,
\begin{eqnarray} 
&\mathcal{L}_1&  \supset  {\kappa_{\gamma\gamma}}\phi^{\ast}\phi A^{\mu\nu}A_{\mu\nu}+{\kappa_{\gamma Z}}\phi^{\ast}\phi A^{\mu\nu}Z_{\mu\nu} 
+{\kappa_{ZZ}}\phi^{\ast}\phi Z^{\mu\nu}Z_{\mu\nu}
+{\kappa_{WW}}\phi^{\ast}\phi W^{\mu\nu}W_{\mu\nu}~~~~~~~~~~~~~~(D=6),\label{eq:gammaz0}\\
&\mathcal{L}_2&  \supset  {\kappa_{\gamma\gamma}}\bar{\chi}\chi A^{\mu\nu}A_{\mu\nu}+{\kappa_{\gamma Z}}\bar{\chi}\chi A^{\mu\nu}Z_{\mu\nu} 
+{\kappa_{ZZ}}\bar{\chi}\chi Z^{\mu\nu}Z_{\mu\nu}
+{\kappa_{WW}}\bar{\chi}\chi W^{\mu\nu}W_{\mu\nu}~~~~~~~~~~~~~~~~~~~(D=7),\label{eq:gammaz1}\\
&\mathcal{L}_3&  \supset  {\kappa_{\gamma\gamma}}\bar{\chi}i\gamma_5\chi A^{\mu\nu}\tilde{A}_{\mu\nu}+{\kappa_{\gamma Z}}\bar{\chi}i\gamma_5\chi A^{\mu\nu}\tilde{Z}_{\mu\nu} 
+{\kappa_{ZZ}}\bar{\chi}i\gamma_5\chi Z^{\mu\nu}\tilde{Z}_{\mu\nu}
+{\kappa_{WW}}\bar{\chi}i\gamma_5\chi W^{\mu\nu}\tilde{W}_{\mu\nu}~~~(D=7).\label{eq:gammaz2}
\end{eqnarray}
Here the physical $W$ denote only the charged components. While $\kappa_{WW} = \kappa_{2}$, the other coefficients are related by the rotation of Weinberg angle,
\begin{eqnarray}
 \kappa_{\gamma \gamma } & = & \kappa_1 \cos^{2} \theta_W + \kappa_2 \sin^{2} \theta_W \, , \\  
 \kappa_{Z Z } & = &  \kappa_2 \cos^{2} \theta_W + \kappa_1 \sin^{2} \theta_W \, , \\
 \kappa_{Z \gamma } & = &  (\kappa_2 - \kappa_1) \sin 2 \theta_W \, . 
\end{eqnarray}
The $\kappa$ coefficients are dimensionful and we will denote \textit{$\Lambda^{-2}_{VV} (D=6)$} or $\Lambda^{-3}_{VV} (D=7) \equiv \kappa_{VV}, V=\gamma,Z$ for the convenience of notation. Generally $\Lambda$ absorbs the couplings and its explicit form in complete UV models would be consist of both SM and new physics couplings and/or masses scales. As an example case, the singlet fermionic DM $\chi$ couples two intermediate states~\cite{Weiner:2012gm}: a fermion $\psi$ and a scalar $\varphi$, which are charged under $SU(2)_L\times U(1)_Y$ and heavier than DM $\chi$. The effective diboson interaction can be generated by the $\psi$ and $\varphi$ loop. The corresponding energy scale $\Lambda$ is then given as $\Lambda^{-3}=g^2\lambda^2C_\psi\mathcal{F}/(48 \pi^2 M^3_\psi)$, where the lower case $\lambda$ is the coupling between $\chi$, $\psi$ and $\varphi$, $C_\psi$=1/2 and $\mathcal{F}$ is the form factor. Considering natural coupling sizes, the additional coefficients enhances $\Lambda$ by one order of magnitude in comparison to $\psi$ and $\varphi$ masses. 

Admittedly in high energy collision processes at colliders, a $\Lambda$ comparable or lower than the center of mass energy can lead to theoretical issues with the effective operators, that the heavy states become accessible at such energies, causing large corrections to the effective operator approach. This brings significant uncertainty to the accuracy of probing the effective operator's scale especially in case of a limited luminosity that does not constrain $\Lambda$ to higher scales than the collision energy. Simplified models with explicit vertexes to heavy states, that fully account for the production of accessible heavy intermediate particles are also popular in current collider searches~\cite{Abdallah:2015ter, Abercrombie:2015wmb, Boveia:2016mrp, Albert:2017onk}. Nevertheless, for specific simplified models, the collider constraint becomes very model dependent and involve a larger number of model parameters than a simple $\Lambda$. Here we use effective operator approach, and consider the CEPC's sensitivity on the effective $\Lambda$ as the lowest order yet a direct estimate of the $Z$-pole runs' capability of testing a diboson operator's energy scale. As we will demonstrate in Section~\ref{sect:cepc}, the $Z$-pole sensitivity for the dimension-6 operator can achieve to be much higher than the center-of-mass energy.

\begin{figure}[h]
  \centering
  \includegraphics[width=0.40\textwidth]{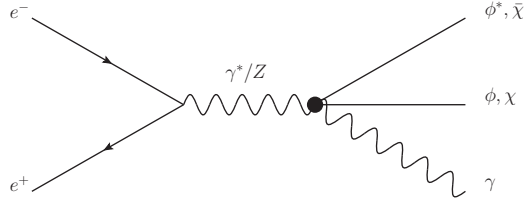}%
  \caption{DM production processes $e^+e^-  \to \phi^{\ast}\phi\gamma, \bar{\chi}\chi\gamma$ with mono-$\gamma$  channel. The $s-$channel $Z$ will be on-shell during $Z$-pole runs at the CEPC.}
  \label{fig:eE_AxX}
\end{figure}

\section{Semi-visible Z boson decay}
\label{sect:z_spec}

The $e^+e^-$ collision is mostly sensitive to the $\gamma,Z$ terms in Eq.~\ref{eq:gammaz0}-\ref{eq:gammaz2}. In comparison, probing the effective $WW$ couplings requires at least one more weak interaction vertex, and is less constrained.

\begin{figure}[h]
  \centering
  \includegraphics[width=0.40\textwidth]{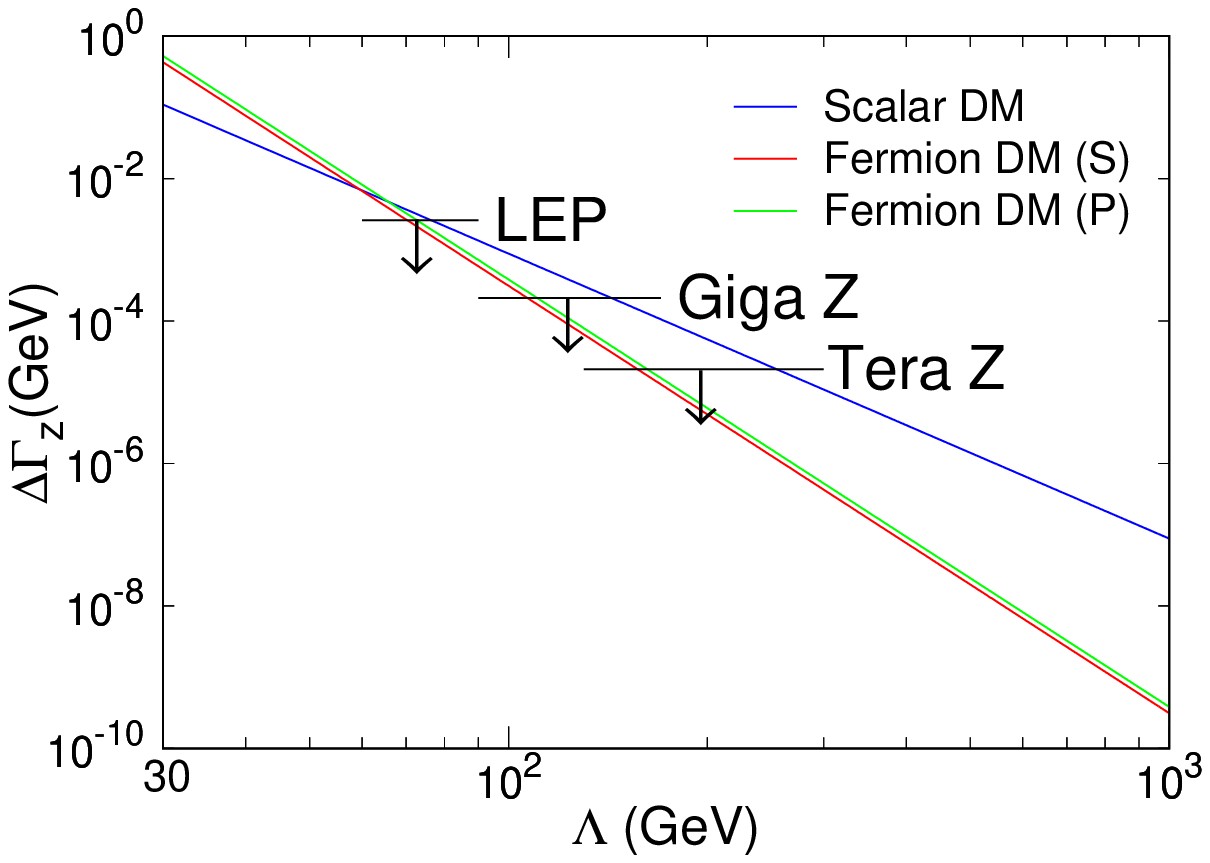}%
      \hspace{0.08\textwidth}%
  \includegraphics[width=0.40\textwidth]{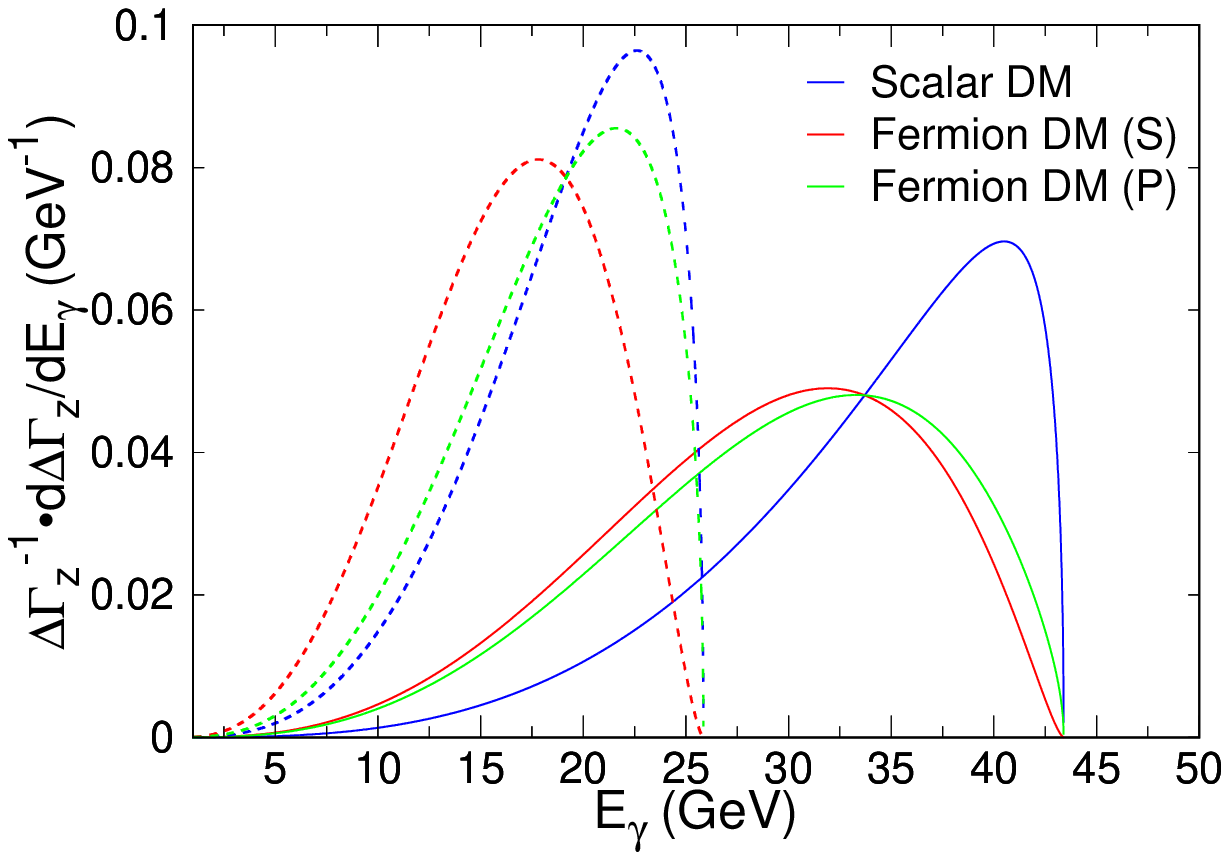}
  \caption{$\Delta\Gamma_Z$ dependence on $\Lambda_{\gamma Z}$(left) at $m_{\phi, \chi}$=10 GeV and the normalized photon energy spectrum $\Delta\Gamma^{-1}_Z \cdot {\rm d}\Delta\Gamma_Z/{\rm d}E_{\gamma}$ (right) for light DM masses with $\Lambda_{\gamma Z}$ fixed at 200 GeV. The blue, red, green (dash) line
denote scalar DM and fermion DM with scalar(S) and pseudo-scalar(P) types, respectively. The solid and dash lines denote DM mass at 10 and 30 GeV, respectively. The black lines (left) denote $\Delta\Gamma_Z=2.6$ (LEP), $2.1\times10^{-1}$ (Giga Z) and $2.1\times 10^{-2}$ (Tera Z) MeV. The latter two are estimated by scaling from projected CEPC luminosities, as $\Delta\Gamma_{{\rm CEPC}}/\Delta\Gamma_{{\rm LEP}}=(\mathcal{L}_{\rm LEP}/\mathcal{L}_{\rm CEPC})^{1/2}$.}
\label{fig:zwidth}
\end{figure}

For a light DM mass, the effective DM $\gamma Z$ coupling lets the physical $Z$ decay semi-visibly into a photon and a DM pair as illustrated in Fig.~\ref{fig:eE_AxX}. This decay would contribute to the total $Z$ width, as well as to the invisible width due to the partially invisible final state. This three-body decay rate is,
\be
{\rm d}\Gamma=\frac{1}{2M_Z}|\mathcal{M}|^2{\rm d}\Phi_3,
\ee
with
\begin{eqnarray}
|\mathcal{M}|^2&=&\frac{8}{3\Lambda^4}(M_ZE_{\gamma})^2 \hspace{0.5cm} (D=6),\label{eq:matrix_element0}\\
|\mathcal{M}|^2_{\rm S}&=&\frac{16}{3\Lambda^6}(M_ZE_{\gamma})^2(M_Z^2-4m^2_{\chi}-2M_ZE_{\gamma}) \hspace{0.5cm}  (D=7),\label{eq:matrix_element1}\\
|\mathcal{M}|^2_{\rm P}&=&\frac{16}{3\Lambda^6}(M_ZE_{\gamma})^2(M_Z^2-2M_ZE_{\gamma})\hspace{0.5cm}  (D=7),\label{eq:matrix_element2}
\end{eqnarray}
where $\Phi_3$ is the three-body phase-space. The subscript S and P indicate scalar and pseudo-scalar types for the fermionic DM bilinear product. The DM energy $E_{\phi, \chi}$ can be integrated out and Eq.~\ref{eq:matrix_element0}-\ref{eq:matrix_element2} are written in terms of the photon energy $E_{\gamma}$ which is the only visible particle in the final state. The differential width can be written as,
\begin{eqnarray} 
\frac{{\rm d}\Gamma}{{\rm d} {E}_{\gamma}}&=&\frac{M_ZE_{\gamma}^3\sqrt{(M^2_Z-2E_{\gamma}M_Z-2m^2_{\phi})^2-4m^4_{\phi}}}{24\pi^3 \Lambda^4(M^2_Z-2E_{\gamma}M_Z)},\\
\frac{{\rm d}\Gamma_{\rm S}}{{\rm d}{ E}_{\gamma}}&=&\frac{M_ZE_{\gamma}^3(M_Z^2-4m^2_{\chi}-2M_ZE_{\gamma})\sqrt{(M^2_Z-2E_{\gamma}M_Z-2m^2_{\chi})^2-4m^4_{\chi}}}{12\pi^3 \Lambda^6(M^2_Z-2E_{\gamma}M_Z)},\\
\frac{{\rm d}\Gamma_{\rm P}}{{\rm d}{ E}_{\gamma}}&=&\frac{M_ZE_{\gamma}^3\sqrt{(M^2_Z-2E_{\gamma}M_Z-2m^2_{\chi})^2-4m^4_{\chi}}}{12\pi^3 \Lambda^6},
\end{eqnarray}
with the photon energy range from 0 to $\frac{1}{2M_Z}(M_Z^2-4m^2_{\phi, \chi})$. Note that with the effective vertex there is no infrared divergence and the photon has a hard spectrum that can be readily searched, as shown in Fig~\ref{fig:zwidth}.

The semi-visible contribution to $Z$ width can be a test for $\Lambda_{\gamma Z}$, as illustrated in the left panel in Fig.~\ref{fig:zwidth}. The right panel denotes the normalized photon energy spectrum $\Delta\Gamma^{-1}_Z \cdot {\rm d}\Delta\Gamma_Z/{\rm d}E_{\gamma}$ (right) for light DM masses with $\Lambda_{\gamma Z}$ fixed at 200 GeV. The blue, red, green (dash) line denote scalar DM and fermion DM with scalar(S) and pseudo-scalar(P) type, respectively. The solid and dash lines denote DM mass 10 and 30 GeV. Note that for collider searches, the difference for fermion DM production rate of S and P types in mono-photon channel only emerge at large DM mass and the former is smaller than the latter due to extra $M_\chi$ forms in squared matrix element $|\mathcal{M}|^2$, as shown in Fig.~\ref{fig:zwidth}. The LEP uncertainty on the invisible $Z$ width $\Delta\Gamma_{{\rm inv}}<$2.6 MeV at the integrated luminosity of ${\rm 161pb^{-1}}$~\cite{Abbiendi:2000hu}. As a full $Z$-pole data analysis is beyond the scope of this paper, here we make a simple estimate for the CEPC's invisible width uncertainty based on the design luminosities. With the luminosity of ${\rm 25fb^{-1}}$ (Giga Z) and ${\rm 2.5ab^{-1}}$ (Tera Z)~\cite{CEPC-SPPCStudyGroup:2015esa}, we scale $\Delta\Gamma\propto {\cal L}^{-1/2}$, thus $\Delta\Gamma_{{\rm CEPC}}/\Delta\Gamma_{{\rm LEP}}=(\mathcal{L}_{\rm LEP}/\mathcal{L}_{\rm CEPC})^{1/2}$ and then the projected $\Delta\Gamma$ are $2.1\times 10^{-1}$ and $2.1\times 10^{-2}$ MeV respectively. Note the invisible width measurement is generally subject to uncertainty from multiple $Z$ decay channels. Better sensitivities can be obtained by focusing on the mono-photon channel, as we will discuss in the following section. 


\section{Mono-photon searches}
\label{sect:cepc}

At the $e^+e^-$ collider, the effective DM diboson couplings give rise to mono-$\gamma$ and mono-$Z$ signals. Both channels are sensitive probes due to a clean and identifiable SM background if compared to hadron colliders. The mono-$Z$ photon is favorably tested off $Z$-pole and has been recently studied by Ref.~\cite{Yu:2014ula}. We focus on the monophoton signal that receives on-shell resonance enhancement at the CEPC's $Z$-pole energy. While $\Lambda_{\gamma\gamma}$ also contributes to this process, its contribution is not resonance enhanced. Therefore the $Z$-pole is a good probe $\Lambda_{\gamma Z}$ that is otherwise often subdominant in direct and indirect searches. 

The mono-$\gamma$ process is illustrated in Fig.~\ref{fig:eE_AxX}. The DM is pair-produced in association an energetic photon, which recoils against the invisible DM pair. The photon is not forwardly (beam-line direction) enhanced, hence it leads to large photon transverse momentum ($P_T$) and the recoiling MET, making it a very clean search channel. With a {\cal O(10)} GeV photon $P_T$ cut, the only relevant SM background channel is $e^+e^-\rightarrow \gamma \nu \bar{\nu}$, where the invisible $\nu\bar{\nu}$ splits from a virtual $Z$. As the total energy is capped at $Z$ mass, the virtual $Z\rightarrow\bar{\nu}\nu$ process acquires suppression by a virtuality $\sim E_\gamma$. This background can be efficiently controlled with a $P_T(\gamma)$ cut. While three neutrino flavors contribute equally to the $\gamma\bar{\nu}\nu$ background via virtual $Z$ mediation, $\gamma\bar{\nu_e}\nu_e$ has additional contribution from $t-$channel $W$ exchanges. 

\begin{figure}[h]
  \centering
  \includegraphics[width=0.40\textwidth]{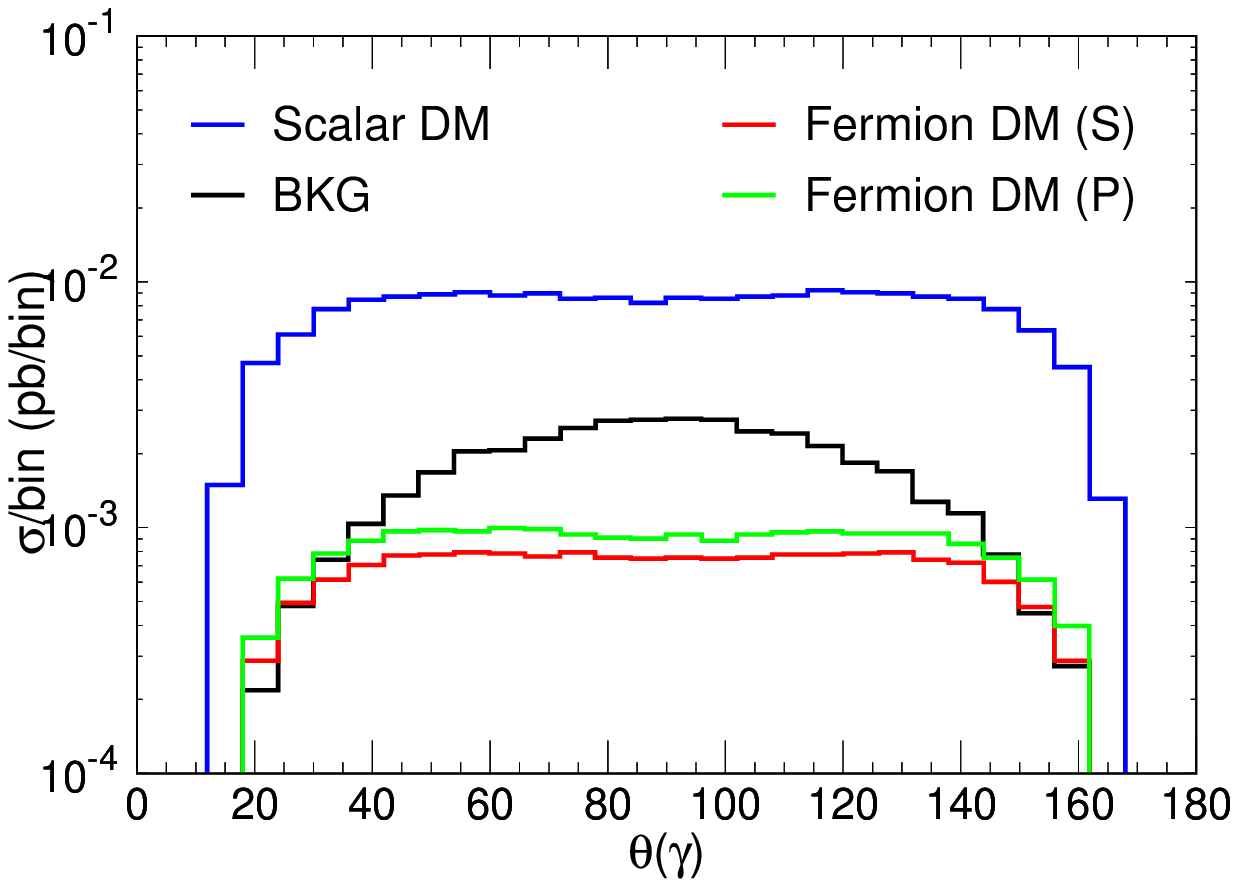}%
    \hspace{0.05\textwidth}%
  \includegraphics[width=0.40\textwidth]{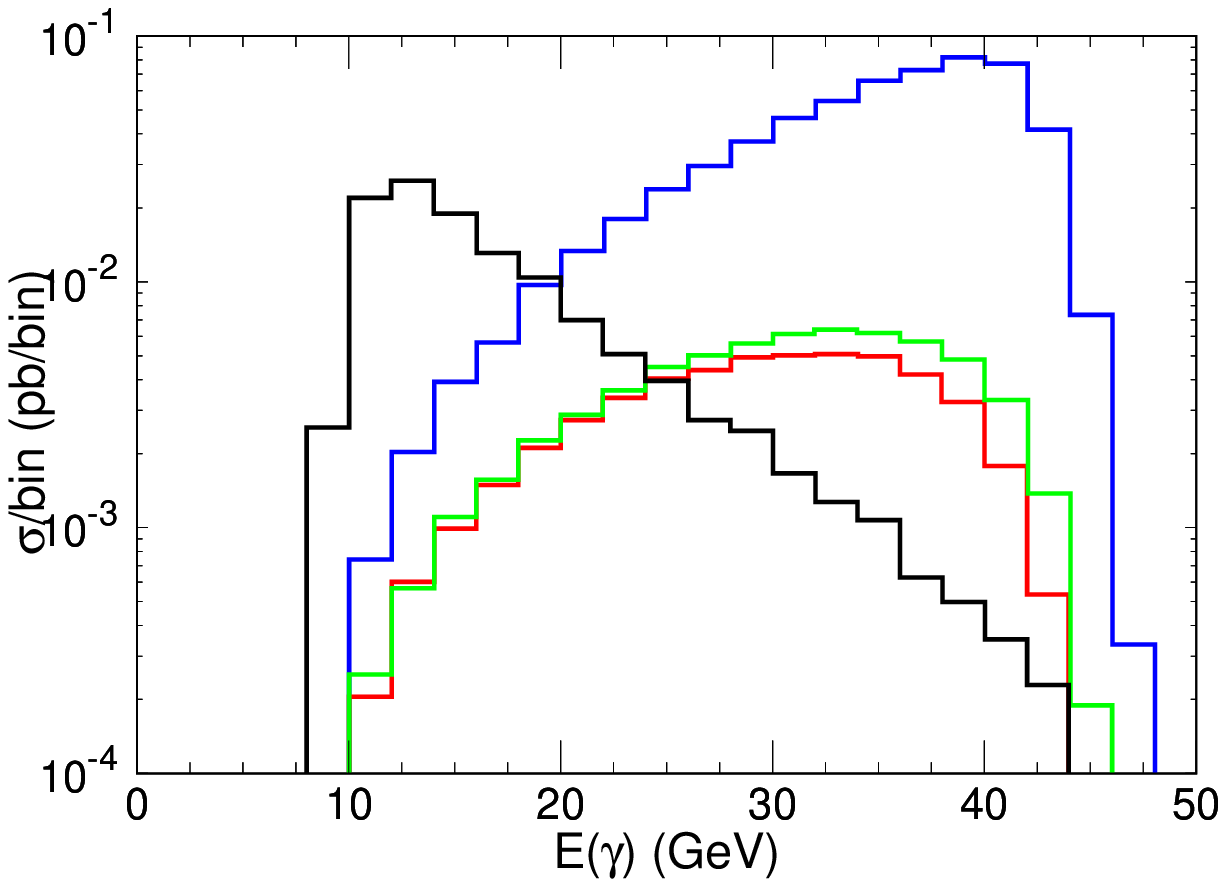}
    \hspace{0.05\textwidth}%
      \includegraphics[width=0.40\textwidth]{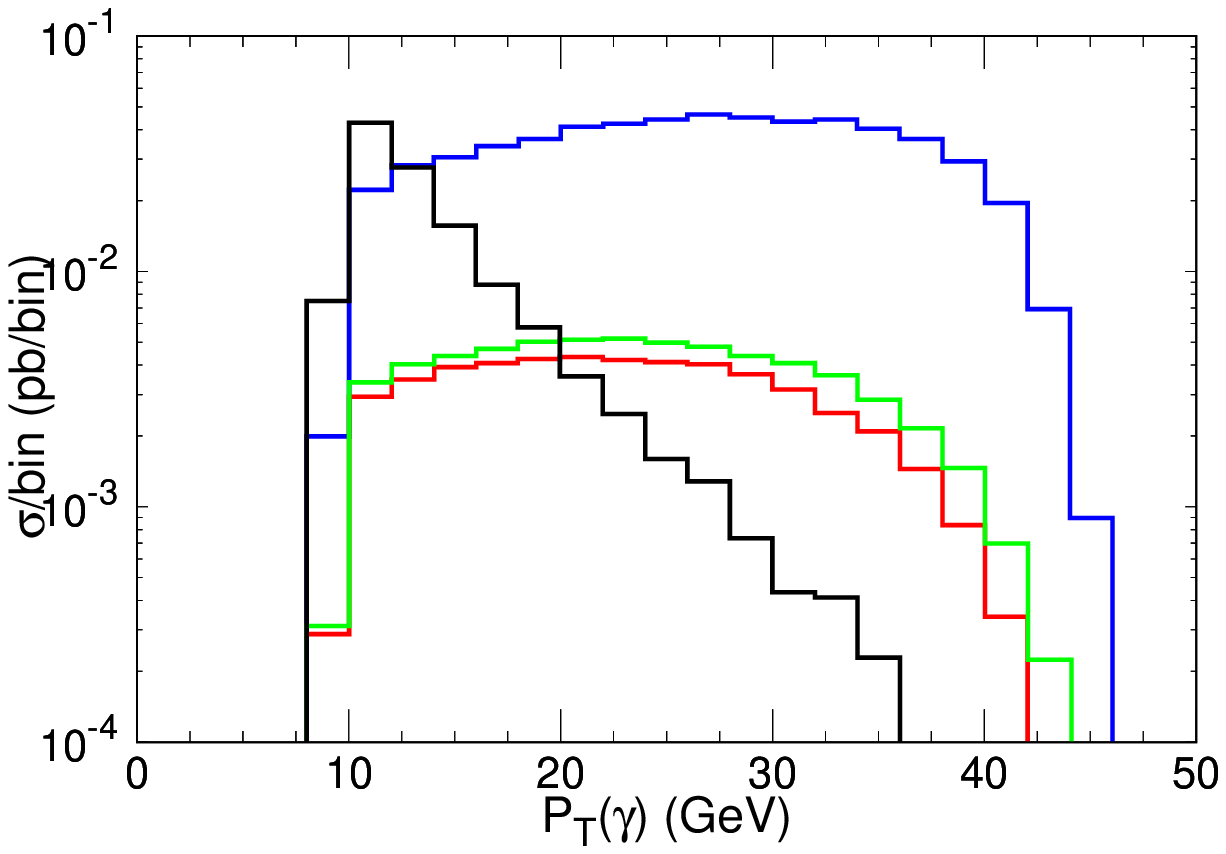}%
    \hspace{0.05\textwidth}%
  \includegraphics[width=0.40\textwidth]{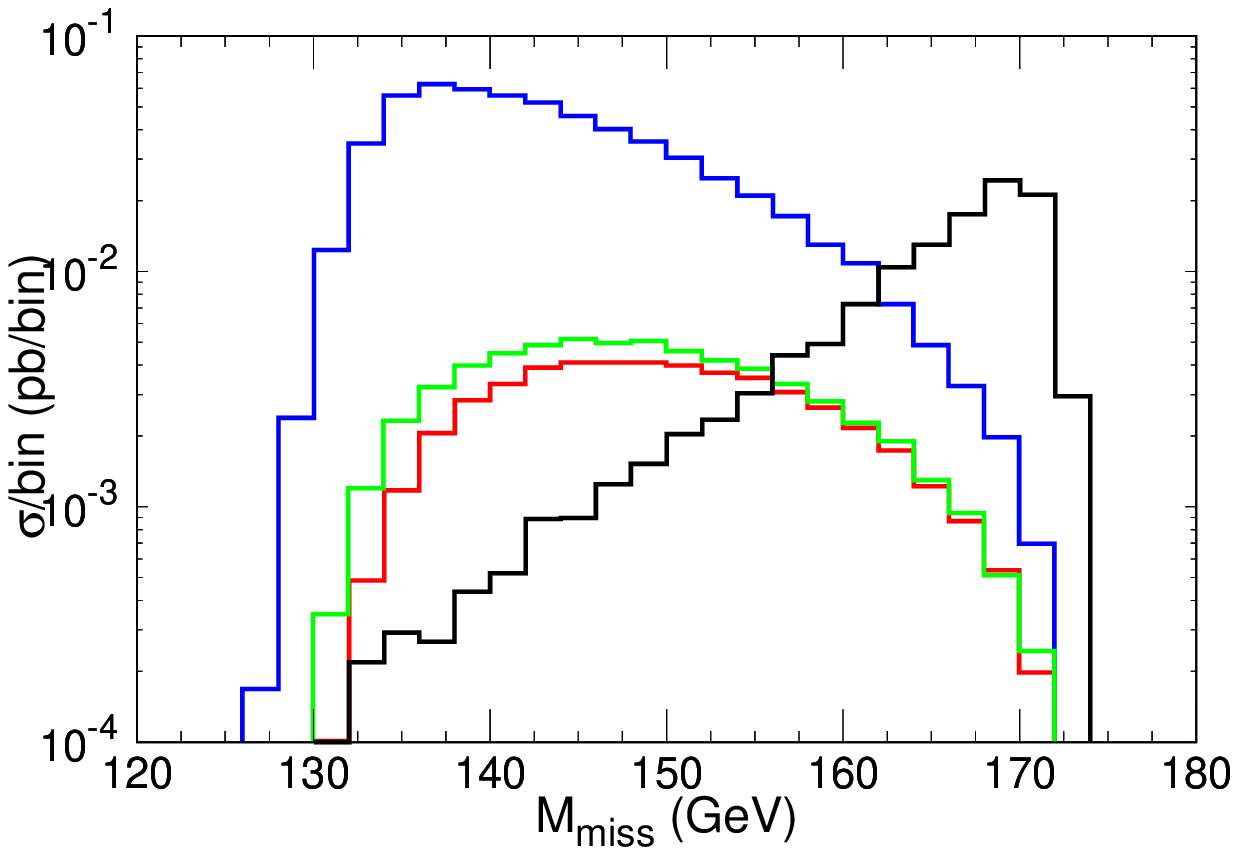}
    \hspace{0.05\textwidth}%
  \caption{Distributions of $\theta(\gamma)$, $E(\gamma)$, $P_T(\gamma)$, $M_{\rm miss}$ at $\sqrt{s}$=91.2 GeV with $\Lambda=200$GeV at the CEPC with $m_{\phi, \chi}$=10 GeV. The background curve (black solid line) is for the irreducible SM $e^+e^-\rightarrow \gamma\bar{\nu}\nu$ process. }
  \label{fig:EP}
\end{figure}

Another background may rise from the soft $e^-e^+$ scattering, that a photon can be emitted and the forward-going $e^\pm$ has a chance of escaping detection if it is still in the high pseudorapidity region. This background is however suppressed by photon $P_T$ and can be very effectively vetoed by a photon $P_T(\gamma)$ cut and non-observation of other detector activity~\cite{Yu:2013aca}, making it much subleading compared to $e^+e^-\rightarrow \gamma\bar{\nu}\nu$ and can be ignored in our analysis.

We use the {\it MadGraph/MadEvent} package~\cite{Alwall:2014hca} to simulate the leading-order signal and background cross-sections at CEPC $Z$-pole runs with basic photon pseudorapidity $\eta$ and $P_T$ cuts. The CEPC detector simulation is done by Delphes~\cite{deFavereau:2013fsa} with the CEPC configurations~\cite{CEPC-SPPCStudyGroup:2015csa}. According to Ref.~\cite{CEPC-SPPCStudyGroup:2015csa}, we adopt the $|\eta(\gamma)|<3$ cut. Then we optimize the $P_T(\gamma)$ cut in our analysis between 25 GeV to 50 GeV to maximize the $S/\sqrt{S+B}$ sensitivity.

In Fig.~\ref{fig:EP}, we show the photon polar angle $\theta$, photon energy $E(\gamma)$, transverse momentum $P_T(\gamma)$ and missing mass $M_{\rm miss}=\sqrt{(p_{e+}+p_{e-}-p(\gamma))^2}$ distributions for $e^+e^-$ center of mass energy at 91.2 GeV. It is clear that at $Z$-pole the signal photons have a broad $P_T$ distribution while the background centers at low $P_T$ and can be distinguished with the photon $P_T(\gamma)$ cut. 


Comparing the fermion and scalar $Z\gamma$ operators in the $e^+e^-\rightarrow \gamma+ \met$ process, the $|\mathcal{M}|^2$ in the fermion case has an extra ${\rm Tr}\left[\slashed{p}_\chi \slashed{p}_{\bar{\chi}}-m^2_\chi\right]=4(p_\chi p_{\bar{\chi}}-m^2_\chi)$ piece that evaluates to $4\sqrt{s}E_{\gamma}$, where $s$ is the collision center of mass energy. This dependence enhances the cross-section of their fermion operators more than the that from the scalar operator, but the fermion operator is also suppressed by $\Lambda$ to one higher order. As a result, for the same $\Lambda$ we find a larger production cross-section by the scalar DM-diboson operators at the $Z$-pole and 240 GeV energy, while for a 500 GeV energy at the ILC the fermion DM operators would yield larger cross-sections.

\begin{table}[h]
\begin{center}
\renewcommand{\arraystretch}{1.2}
\setlength\tabcolsep{0.5em}
\begin{tabular}{cccc|cccc|cccc}
\hline\hline
\multicolumn{4}{c|}{$\sqrt{s}=91.2\GeV$}& \multicolumn{4}{c|}{$\sqrt{s}=240\GeV$}& \multicolumn{4}{c}{$\sqrt{s}=500\GeV$ (ILC)}\\
 Cut& $\nu\bar\nu\gamma$ & $\phi^{\ast}\phi\gamma$ & $\bar{\chi}\chi\gamma$ & 
  Cut& $\nu\bar\nu\gamma$ & $\phi^{\ast}\phi\gamma$ & $\bar{\chi}\chi\gamma$&  
  Cut& $\nu\bar\nu\gamma$ & $\phi^{\ast}\phi\gamma$ & $\bar{\chi}\chi\gamma$\\
$P_T(\gamma)\,(\GeV) $  & $\sigma$\,(fb)  & $\sigma$\,(pb)& $\sigma$\,(fb)  & 
$P_T(\gamma)\,(\GeV) $  & $\sigma$\,(pb) & $\sigma$\,(fb) & $\sigma$\,(fb) &
 $P_T(\gamma)\,(\GeV) $  & $\sigma$\,(fb)  & $\sigma$\,(fb) & $\sigma$\,(pb) \\
\hline
 25                & 7.8    & 1.1 &  67    &  35$^*$        & 1.4  & 118 & 62   & 50               &  636 & 535 & 1.3 \\
 30$^*$         & 2.6    & 0.8 &  40    & 40                & 1.2  & 113 & 58   & 70                &  422 & 503 &  1.2\\
 35$^\dagger$        & 0.7   & 0.5 &  18     & 45$^\dagger$          & 1.1  & 107 & 53   & 90$^*$        &  304 & 462 &  1.0\\
 40                & 0.1   & 0.2 &  4.1    & 50                 & 1.0  & 101 & 48   & 110$^\dagger$        &  228 & 413 & 0.8\\
\hline\hline
\end{tabular}
\end{center}
\caption{Cross sections of SM background and signal processes at $\sqrt{s}=91.2$ and 240 GeV CEPC. Here we also list the $\sqrt{s}=500$ GeV ILC runs' result for the convenience of comparison. Photon pseudorapidity restrict to the central region, $|\eta|<3$ at CEPC and polar angle is $10^\circ<\theta_{\gamma}<170^\circ$ at ILC. For the listed signal cross-sections, the DM mass is fixed at $m_{\phi, \chi}=1\, \GeV$ and $\Lambda_{\gamma Z}=\Lambda_{\gamma\gamma}=200\,\GeV$. The $P_T(\gamma)$ cut with $^\dagger$ or $^*$ is the optimized value for scalar or fermion DM with low mass.}
\label{tab:cuts}
\end{table}

Table~\ref{tab:cuts} lists the signal and background cross sections $\sigma$ after a set of $P_T({\gamma})$ cut values from 25 to 50 GeV at CEPC and from 50 to 110 GeV at ILC. The show-case signal cross-sections assume a light $\phi, \chi$ mass at 1 GeV and $\Lambda_{\gamma Z}=\Lambda_{\gamma\gamma}=200$GeV. At $Z$-pole, $\Lambda_{\gamma Z}$ contribution dominates. For heavier $m_{\phi, \chi}$, the final state photon energy is kinematically limited and becomes softer, leading to larger SM background and lower sensitivity to $\Lambda$. This photon would eventually vanish as $m_{\phi, \chi}$ approaches to the beam energy, as illustrated in Fig.~\ref{fig:EP}. The superscript $^\dagger$ and $^*$ denote optimized $P_T(\gamma)$ cut for scalar and fermion DM at $Z$-pole and 240 GeV (CEPC), as well as 500 GeV (ILC) runs in the low DM mass limit, respectively. To obtain the best experimental sensitivity we also considered the LEP angle cut: $20^\circ<\theta<35^\circ, 45^\circ<\theta<135^\circ, 145^\circ<\theta<160^\circ$~\cite{Acciarri:1997im} and a missing mass cut: $M_{\rm miss}<140$ GeV. The relevant distributions are illustrated in Fig.~\ref{fig:EP}. By applying these cuts after $P_T(\gamma)$ and $\eta$ cuts, the photon angle cut will not further improve the sensitivity, and the $M_{\rm miss}$ cut only gives ${\cal O}(10^{-2})$ corrections. Therefore we consider the photon $P_T$ cut sufficient for this study.

\begin{figure*}[h]
  \centering
  \includegraphics[width=0.33\textwidth]{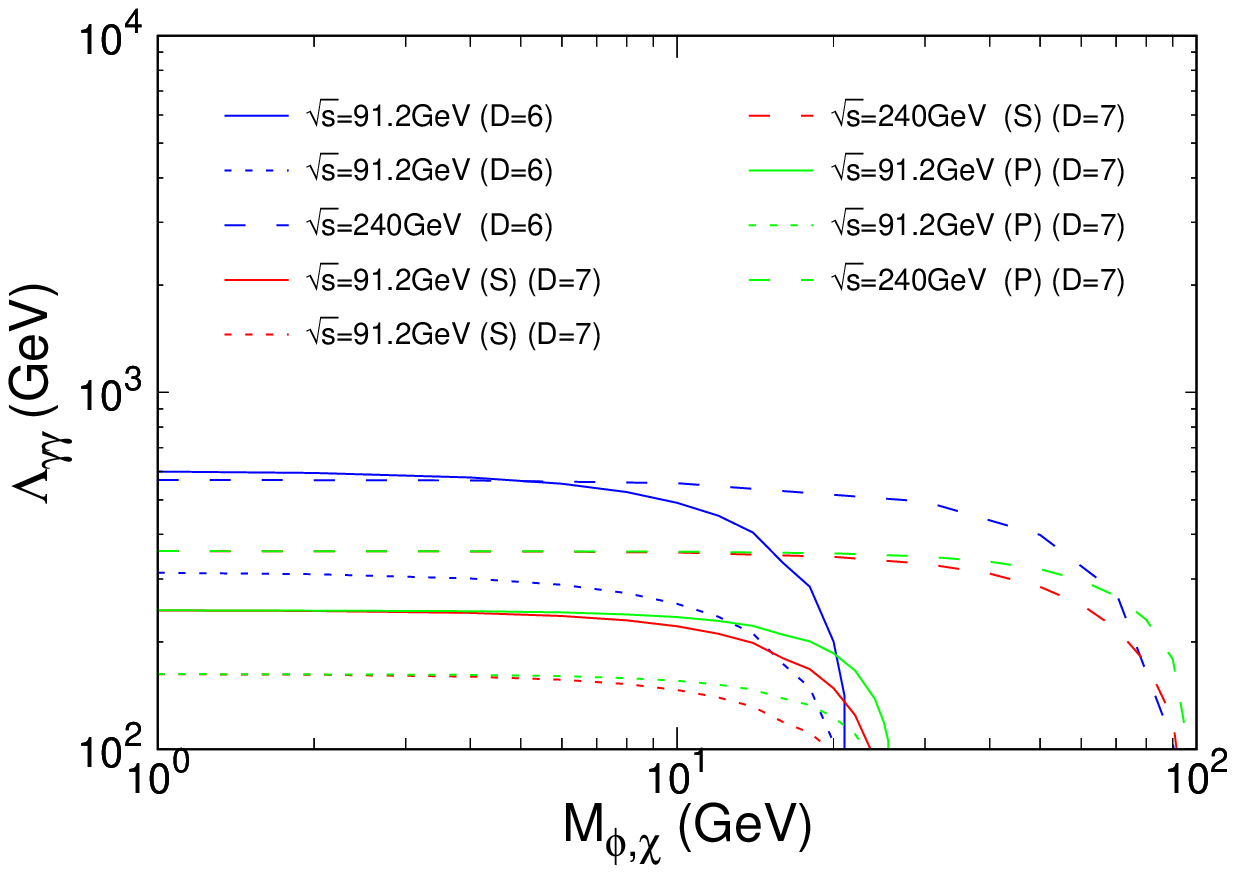}%
 \hspace{0.001\textwidth}%
        \includegraphics[width=0.33\textwidth]{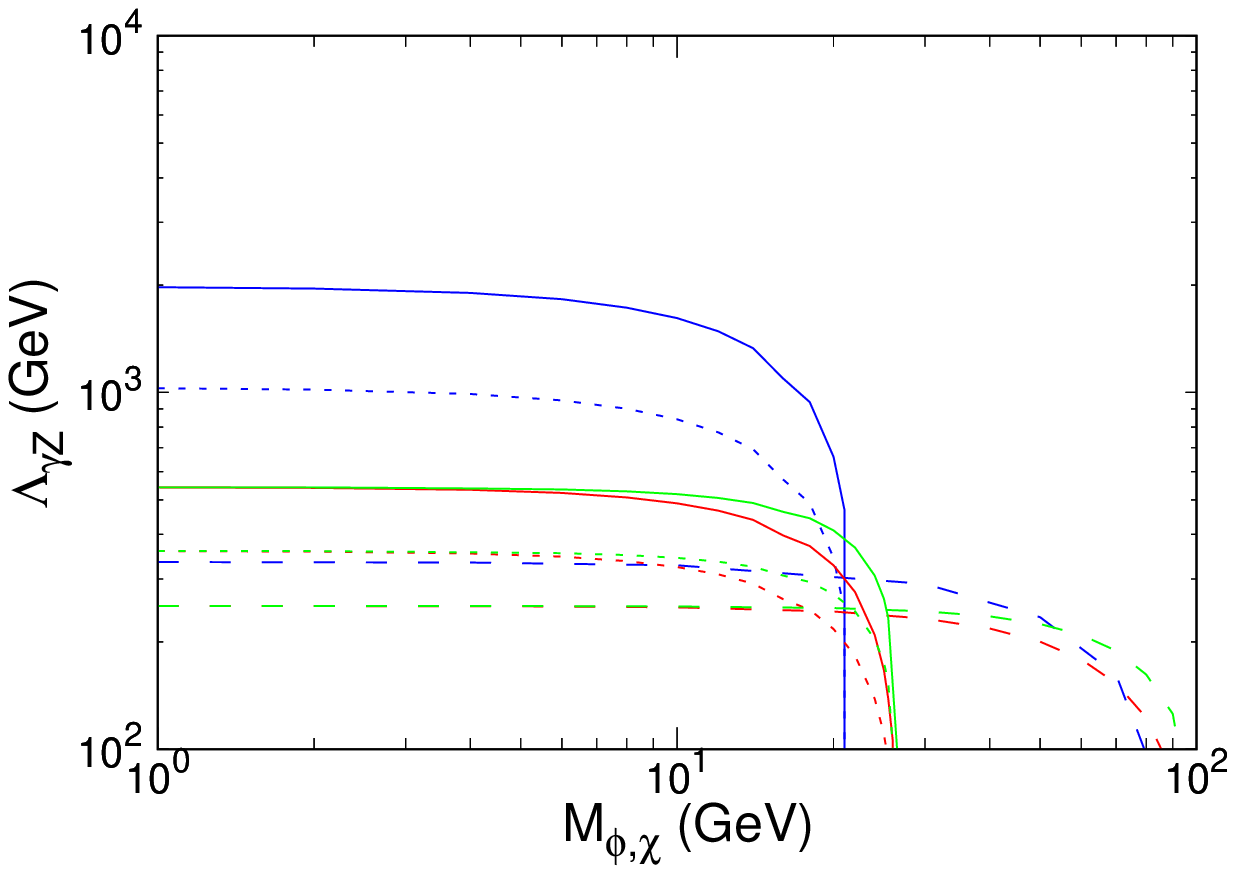}%
    \hspace{0.001\textwidth}%
  \includegraphics[width=0.33\textwidth]{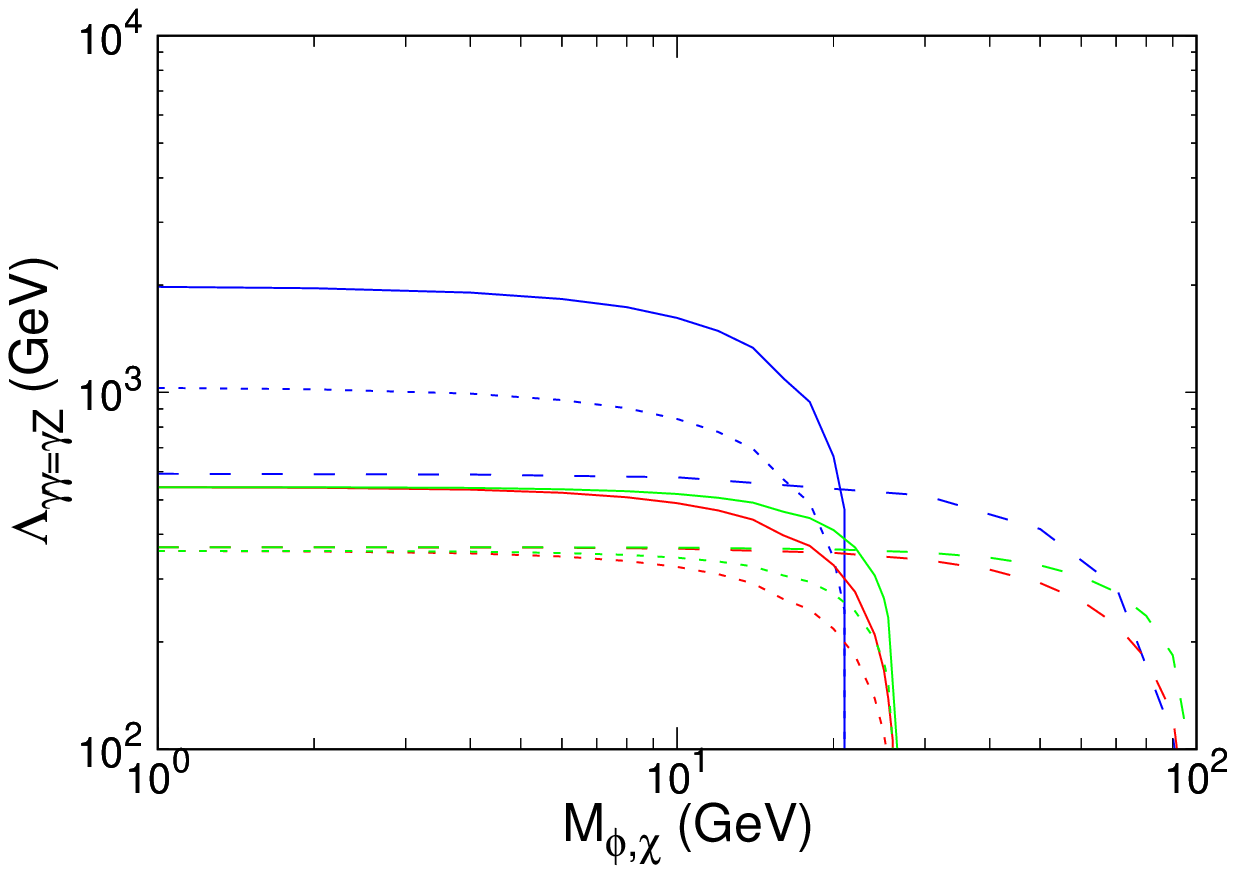}
  \caption{The $3\sigma$ reaches with unpolarized $e^\pm$ beams in the $m_{\phi, \chi}-\Lambda$ plane. The blue, red and green solid/dotted/dashed lines denote scalar DM, fermion DM with scalar and pseudo-scalar types, respectively, where the solid, dotted and dashed lines denote $\sqrt{s}$= 91.2GeV at 2.5 ab$^{-1}$, 25 fb$^{-1}$ and 240 GeV at 5 ab$^{-1}$. The left, middle and right panels assume $\Lambda_{\gamma\gamma}$-only, $\Lambda_{\gamma Z}$-only and $\Lambda_{\gamma Z}$=$\Lambda_{\gamma\gamma}$ cases. $P_T(\gamma)$ cut is at 35 (45) GeV for 91.2 (240) GeV runs for scalar DM, $P_T(\gamma)$ cut is at 30 (35) GeV for 91.2 (240) GeV runs for fermion DM that optimizes the sensitivity for a low $m_{\phi, \chi}$.}
\label{fig:Lambda}
\end{figure*}

The design luminosity at the CEPC~\cite{CEPC-SPPCStudyGroup:2015esa} is 25 fb$^{-1}$ (giga-Z) and 2.5 ab$^{-1}$(tera-Z) at the $Z$-pole, and 5 ab$^{-1}$ in the high-energy 240 GeV run. We set $3\sigma$ sensitivity on $\Lambda_{\gamma Z},\Lambda_{\gamma\gamma}$ by requiring $S/\sqrt{S+B}=3$ significance at the specified luminosities. $S, B$ are the event numbers for signal and SM background channels, respectively. The result for prospective $\Lambda_{\gamma Z, \gamma\gamma}$ sensitivities are shown in Fig.~\ref{fig:Lambda} and later in Fig.~\ref{fig:Lambda_gammaz}. The proposed 25 fb$^{-1}$ (2.5 ab$^{-1}$) $Z$-pole luminosity runs can probe $\Lambda_{\gamma Z}$ to 1030 (1970) GeV for scalar DM, to 360 (540) GeV for fermion DM. At 240 GeV, a better sensitivity in $\Lambda_{\gamma\gamma}$ is obtained that a 5 ab$^{-1}$ luminosity can be probe to 590 (360) GeV for scalar (fermion) DM. 

The sensitivity may be further improved by polarized $e^\pm$ beams. A polarized electron source has been discussed in the current CEPC design~\cite{CEPC-SPPCStudyGroup:2015esa}. Here we consider a $\{P_{e^-},P_{e^+}\}=\{80\%,30\%\}$ beam polarization in $e^\pm$ helicity similar to that of the ILC design~\cite{Behnke:2013xla}. The beam polarization $P_{e^\pm}>0$ is right-handed and $P_{e^\pm}<0$ is left-handed. Since the $Z$ boson coupling is larger to the left-handed chiral current of the electron, a left-handed $\{-,+\}$ configuration will lead to higher $Z$ luminosity than that from a right-handed $\{+,-\}$ polarization configuration, and more stringent limits on $\Lambda$. Similarly the SM backgrounds also increase proportionally for a left-handed configured beam polarization. Adopting the $\{-80\%,+30\%\}$ beam polarization, we found the constraint on $\Lambda$ for scalar (fermion) DM operators at $Z$-pole and 240 GeV and can be enhanced by 1.2$\%$ (1.3$\%$) at 2.5 ab$^{-1}$ luminosity, and 11.2$\%$ (7.3$\%$) at 5 ab$^{-1}$ luminosity.



\section{Direct and indirect limits}
\label{sect:limits}

\begin{figure}[h]
  \centering
   \subfigure[Direct detection]{
      \label{subfig:fa} 
   \includegraphics[width=0.25\textwidth]{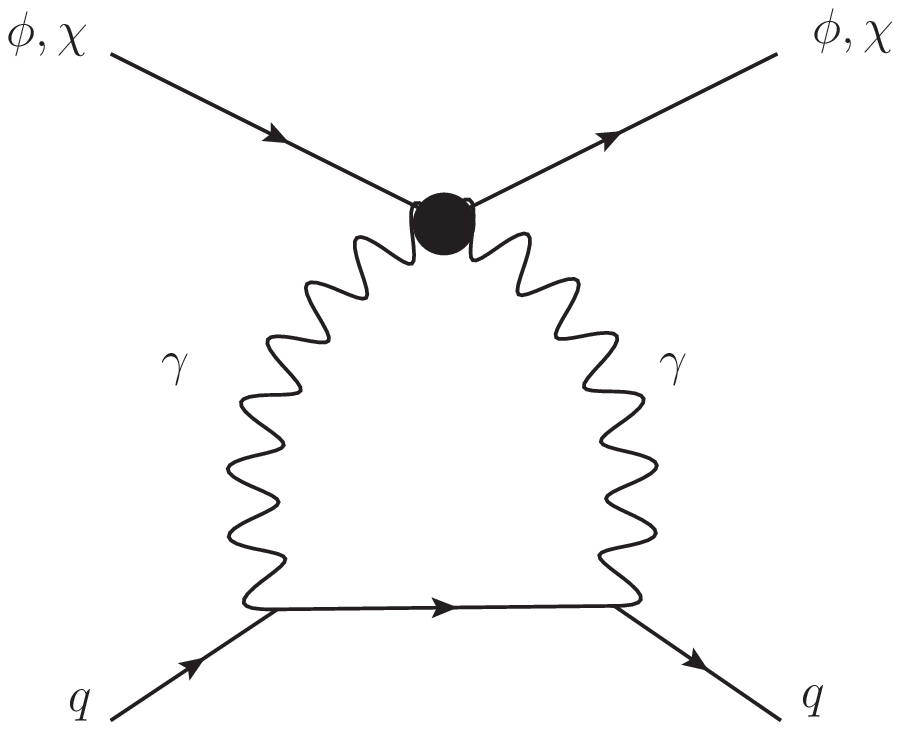}}%
  \hspace{0.01\textwidth} 
  \subfigure[Indirect detection]{ 
     \label{subfig:fb} 
    \includegraphics[width=0.32\textwidth]{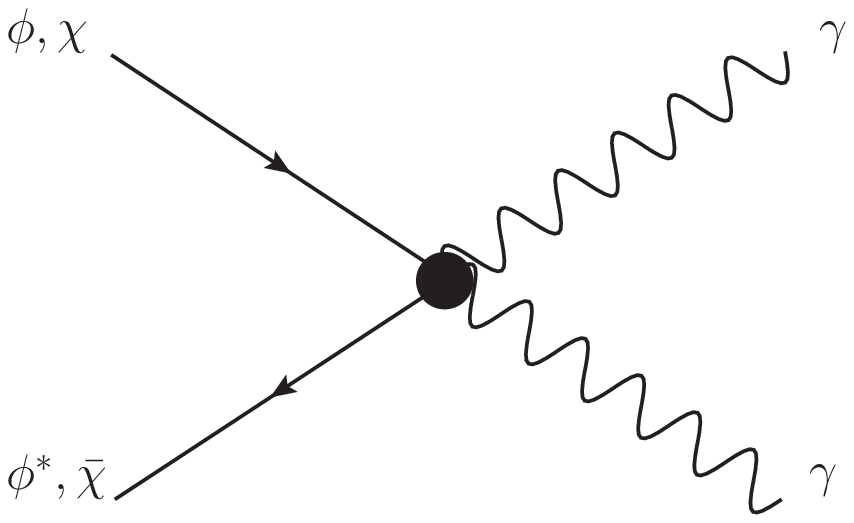}} 
  \caption{Feynman diagrams for DM diphoton scattering off nucleus (left) in direct detection, and DM annihilation process (right) for indirect gamma-ray search. The blob represents the effective diboson-DM vertex.}
  \label{fig:twodiagram}
\end{figure}

In this section we discuss the (mostly) $\Lambda_{\gamma\gamma}$ bounds from current direct and indirect search experiments. The effective diboson interaction allows the DM to scatter off nuclei via a gauge boson loop, as shown in left-panel of Fig.~\ref{fig:twodiagram}. The momentum transfer in direct detection experiments is at keV scale and the diphoton exchange dominates the scattering process, which bears similarity to Rayleigh scattering~\cite{Weiner:2012cb}. Following the procedure in Ref.~\cite{Weiner:2012cb, Frandsen:2012db}, we compute the averaged per nucleon scattering cross-section with,
\begin{eqnarray}
\sigma_{n} &=& \frac{8\mu^2_A\alpha^2Z^4Q^2_0F^2_{{\rm Ray}}(\bar{q})}{\pi^2 m^2_{\phi} A^4 \Lambda_{\gamma\gamma}^4 }\, ~~~(D=6),\label{eq:dd6}\\
\sigma^{\rm S}_{n} &=& \frac{8\mu^2_A\alpha^2Z^4Q^2_0F^2_{{\rm Ray}}(\bar{q})}{\pi^2A^4\Lambda_{\gamma\gamma}^6}~~~~ (D=7),\label{eq:dd7}
\end{eqnarray}
where $A$ is the isotope-averaged number of nucleons, $\mu_A=m_Am_{\phi, \chi}/(m_A+m_{\phi, \chi})$ is the reduced mass (see Appendix A), the charge form factor $F_{{\rm Ray}}(\bar{q})$ drops with rising momentum transfer. The nuclear coherence scale $Q_0\simeq 0.48(0.3+0.89A^{1/3})^{-1}$GeV. Scattering for the pseudo-scalar type for fermion DM is suppressed~\cite{Frandsen:2012db} and we do not discuss it here. The photon mediated scattering is enhanced by the nucleus' number of protons as $Z^4$. The $\gamma Z$-loop contribution is subleading due to $M_Z$ suppression in the heavy propagator. While the $\gamma\gamma,\gamma Z$ interference diagrams can be relevant for $\Lambda_{\gamma Z}\ll \Lambda_{\gamma\gamma}$, the $\gamma Z$ scattering calculation is currently unavailable and is important for future research. Here we only include $\Lambda_{\gamma\gamma}$ contribution in direct detection limits.

\begin{figure} 
  \centering 
  \subfigure[Direct detection]{ 
    \label{subfig:la} 
    \includegraphics[width=0.40\textwidth]{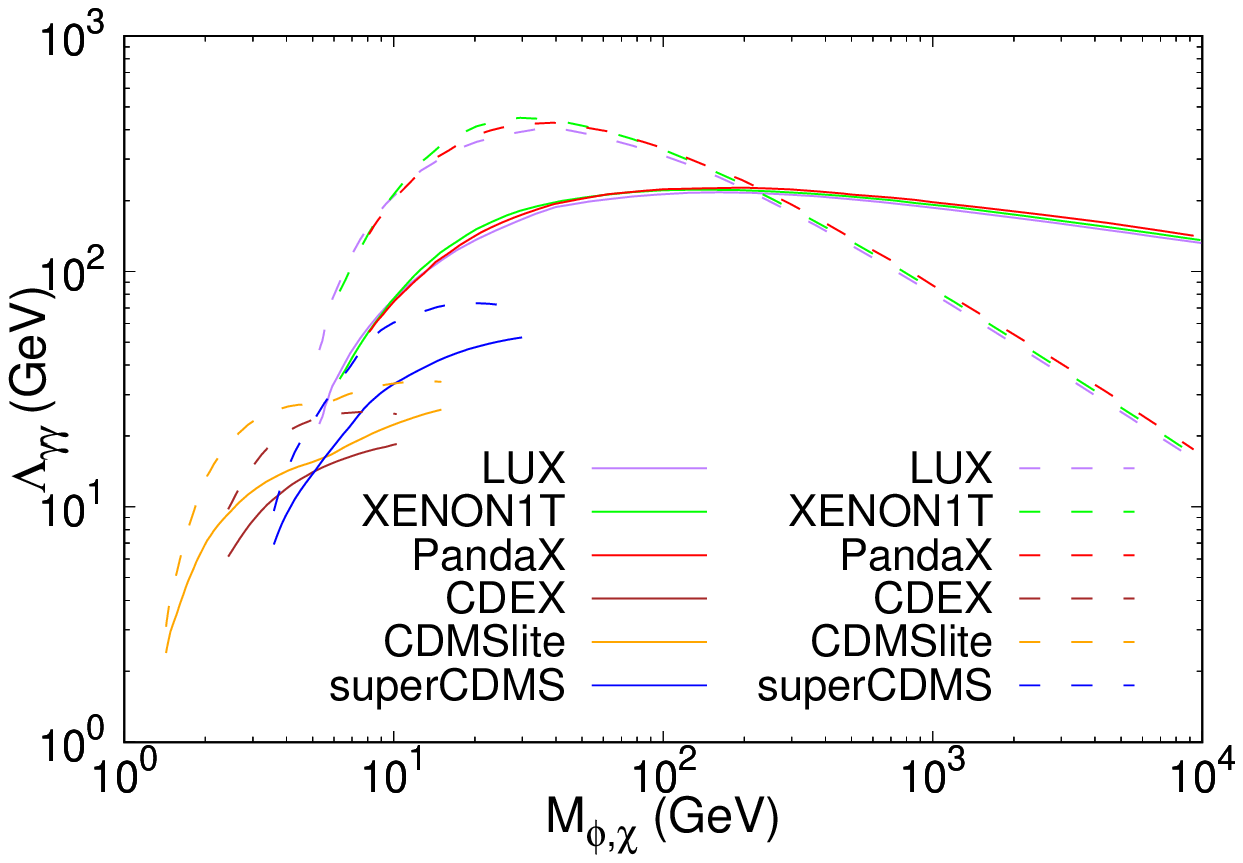}} 
  \hspace{0.01\textwidth} 
  \subfigure[Indirect detection]{ 
    \label{subfig:lb} 
    \includegraphics[width=0.40\textwidth]{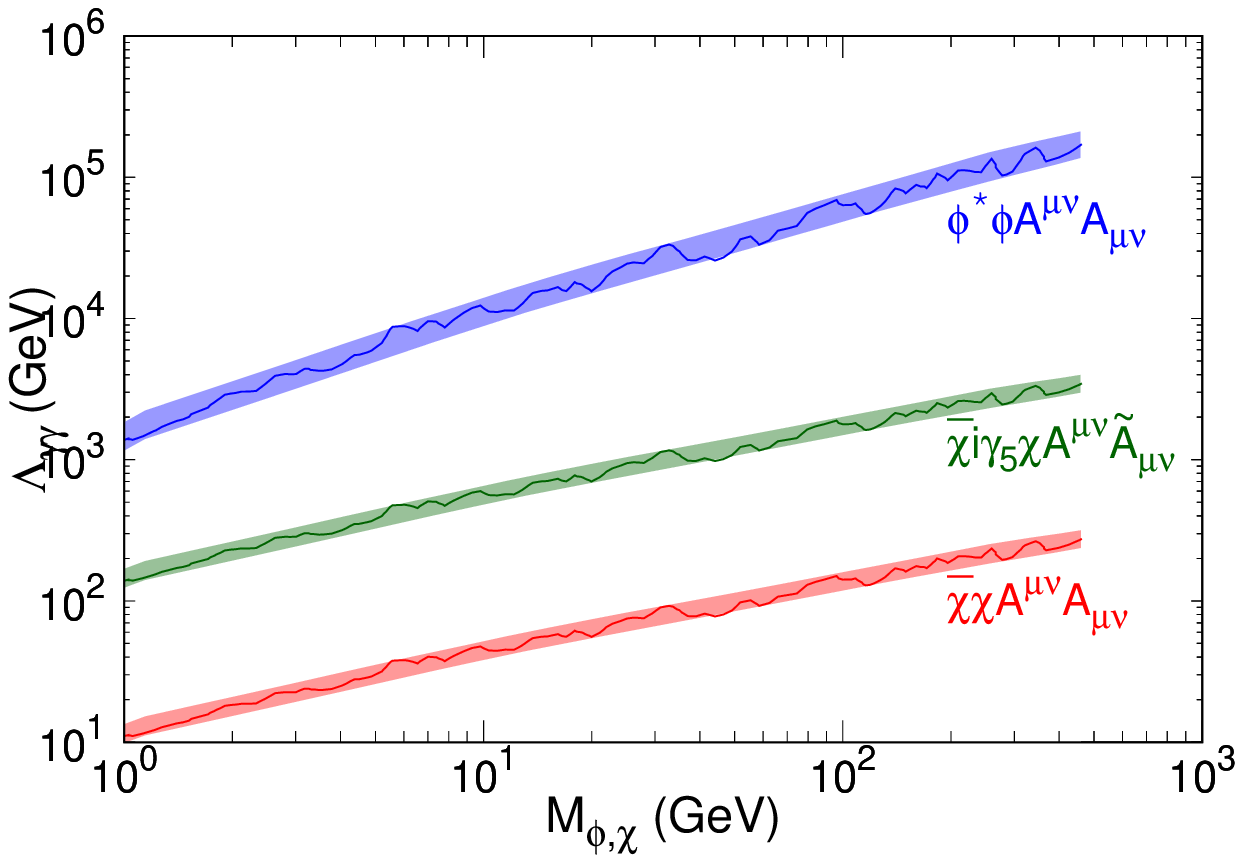}} 
  \caption{Limit on $\Lambda_{\gamma\gamma}$ for $m_{\phi, \chi}$ from direct (left) and indirect (right) detection searches. Direct detection includes latest results from SuperCDMS~\cite{Agnese:2014aze}, CDEX~\cite{Jiang:2018pic}, CDMSlite~\cite{Agnese:2015nto}, XENON1T~\cite{Aprile:2017iyp}, LUX~\cite{Akerib:2016vxi} and PandaX~\cite{Cui:2017nnn} experiments. Here solid(dashed) line denotes fermion with scalar type (scalar) DM for direct detection. Indirect detection uses the Fermi-LAT's monochromatic photon constraint from R3 at 95$\%$ C.L.~\cite{Ackermann:2015lka}  observation region.} 
  \label{fig:Lambda_mchi} 
\end{figure}

A number of existing direction experiments set limits on accessible $m_{\phi, \chi}$ at the CEPC. We illustrate the constraints from a list of recent direct detections results on $\Lambda_{\gamma\gamma}$ in Fig.~\ref{subfig:la}. For $m_{\phi, \chi}>10$ GeV, the latest Xenon-based experiments readily constrain $\Lambda_{\gamma\gamma}$ limit. A lower $m_{\phi, \chi}<4$ GeV would observe a sub-100 GeV $\Lambda_{\gamma\gamma}$ bound from current direct detection results, and may be more effective searched for in future low-threshold nucleus recoil detectors. Ref.~\cite{Weiner:2012cb, Frandsen:2012db, Crivellin:2014gpa, Ovanesyan:2014fha} discussed DM-diboson operators' contribution to nucleon scattering at one loop level, here we follow their results and give the direct-detection constraints. 

For indirect detection, the non-relativistic cross section of DM annihilation into two photons($\phi^{\ast}\phi, \bar{\chi}\chi\to\gamma\gamma$) is dominated by $\Lambda_{\gamma\gamma}$,
\begin{eqnarray}
\langle \sigma v\rangle_{\gamma\gamma}&=&\frac{2m^2_{\phi}}{\pi \Lambda^4_{\gamma\gamma}}~~~(D=6),\\
\langle \sigma v\rangle^{\rm S}_{\gamma\gamma}&=&\frac{m^4_{\chi}v^2}{\pi \Lambda^6_{\gamma\gamma}}~~~(D=7),\\
\langle \sigma v\rangle^{\rm P}_{\gamma\gamma}&=&\frac{4m^4_{\chi}}{\pi \Lambda^6_{\gamma\gamma}}~~~~(D=7),
\end{eqnarray}
for $m_{\phi, \chi}$ below $M_Z/2$. $\Lambda_{\gamma Z}$ dependence only emerges in a small correction from $Z$ mediation as part of the $\phi^{\ast}\phi, \bar{\chi}\chi\rightarrow \gamma (\gamma^*/Z^*\rightarrow \bar{f}f)$ process, which is suppressed by the $\bar{f}f$ mass for virtual photon mediation and $M_Z$ for virtual $Z$ mediation. As a result, $\Lambda_{\gamma\gamma}$'s contribution also by far dominates over that of $\Lambda_{\gamma Z}$. In Fig.~\ref{subfig:lb} we show the 95$\%$ C.L. $\Lambda_{\gamma\gamma}$ constraint from gamma ray line search at Fermi-LAT~\cite{Ackermann:2015lka}.
Note the operators in Eq.~\ref{eq:l1} leads to a $p$-wave annihilation. In case of $s$-wave annihilation by $\bar{\chi}i\gamma^5\chi A^{\mu\nu} \tilde{A}_{\mu\nu}$ interaction, the galactic velocity suppression $v^2\approx 10^{-6}$ is lifted and the $\Lambda_{\gamma\gamma}$ bound improves by one order of magnitude.

\begin{figure}[h]
  \centering
  \includegraphics[width=0.45\textwidth]{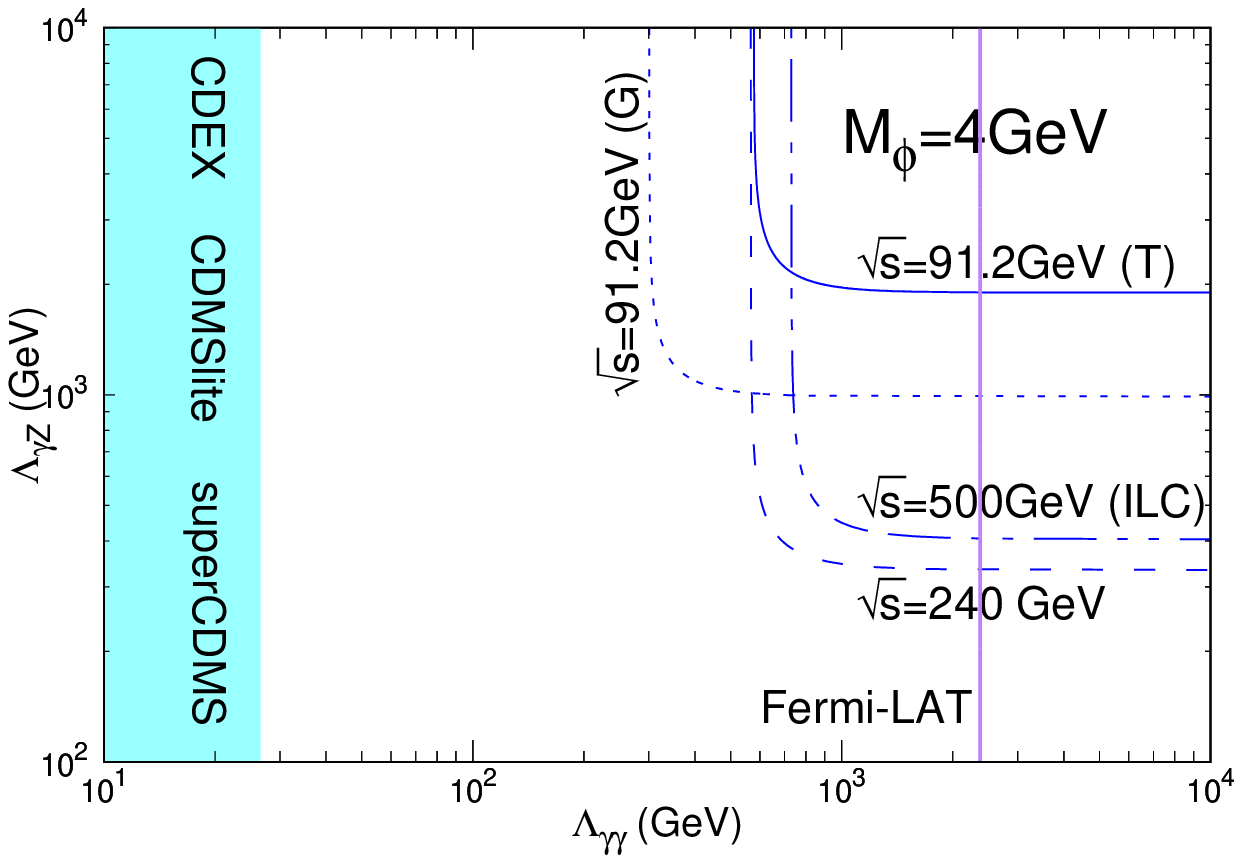}%
  \hspace{0.01\textwidth}%
  \includegraphics[width=0.45\textwidth]{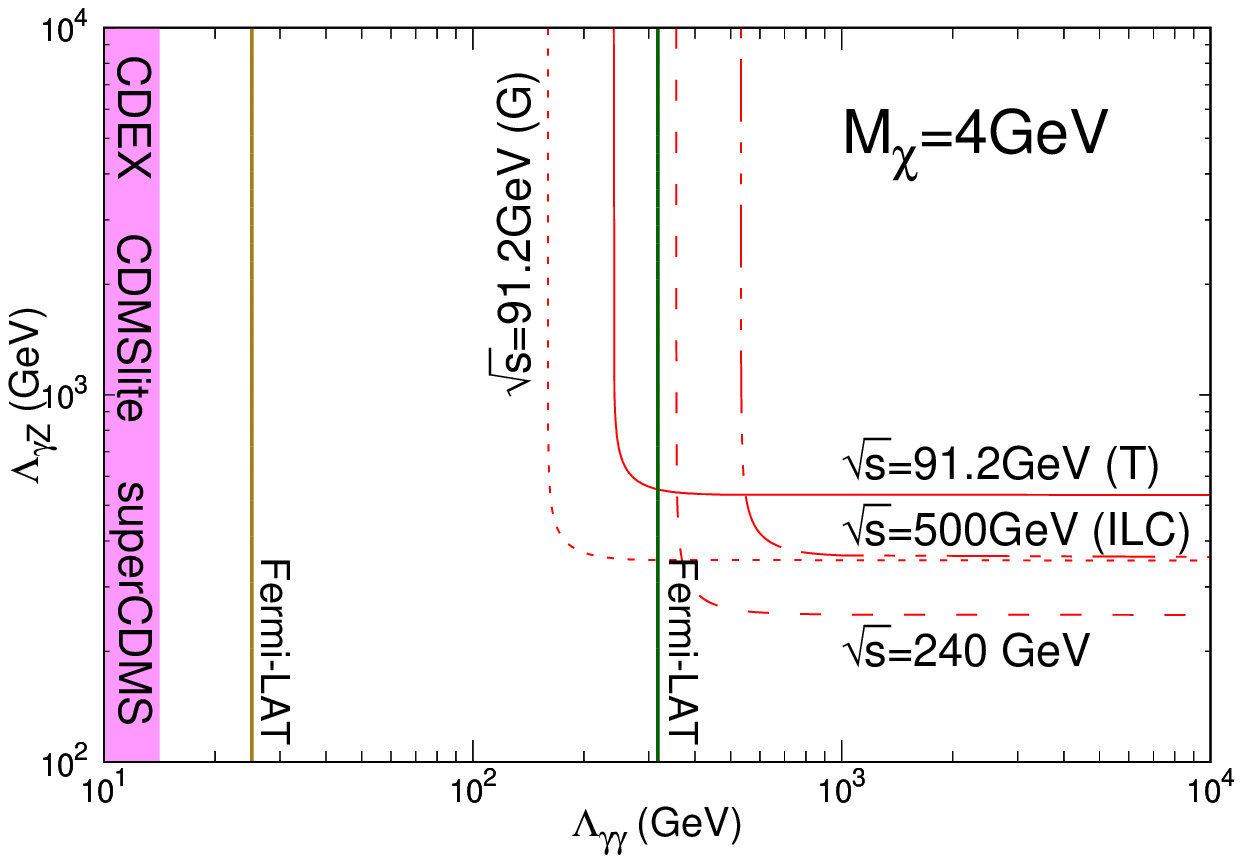}
    \hspace{0.01\textwidth}%
  \includegraphics[width=0.45\textwidth]{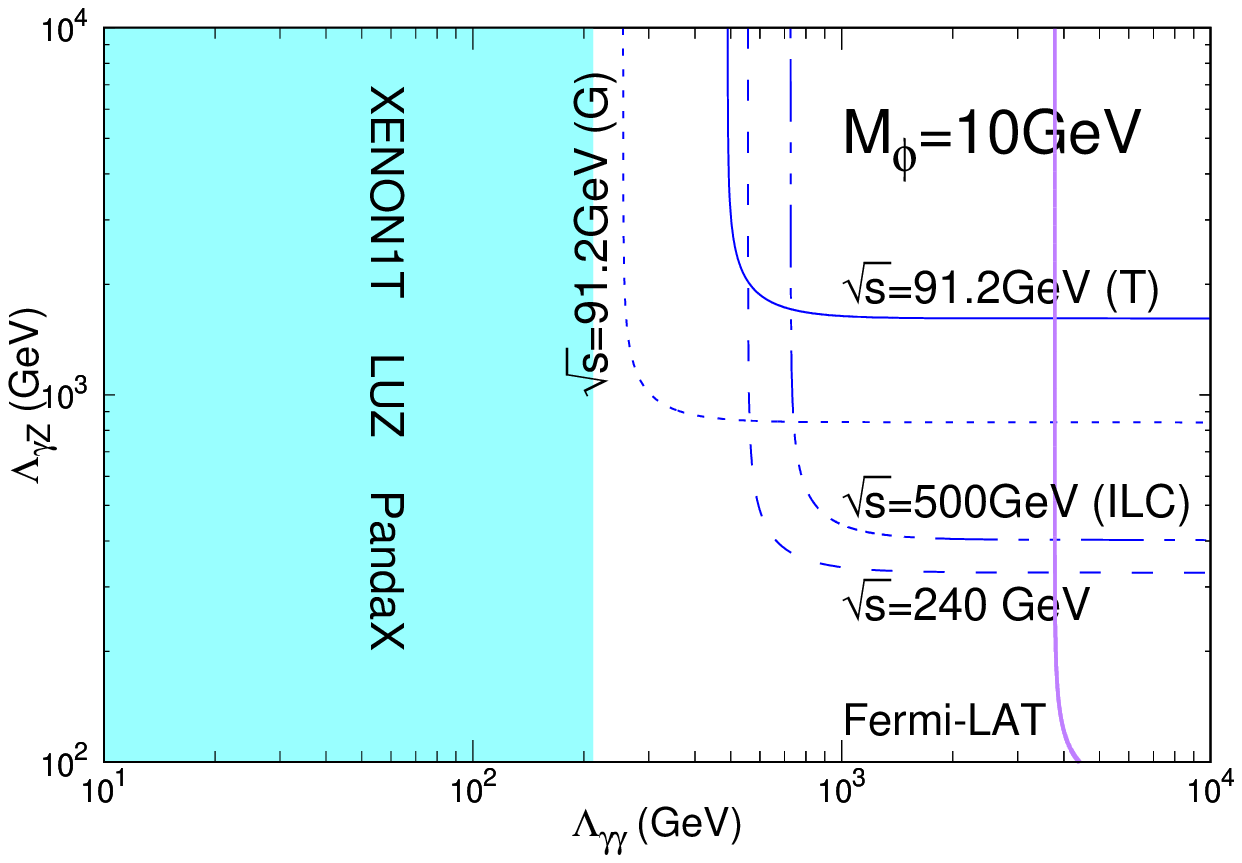}%
    \hspace{0.01\textwidth}%
  \includegraphics[width=0.45\textwidth]{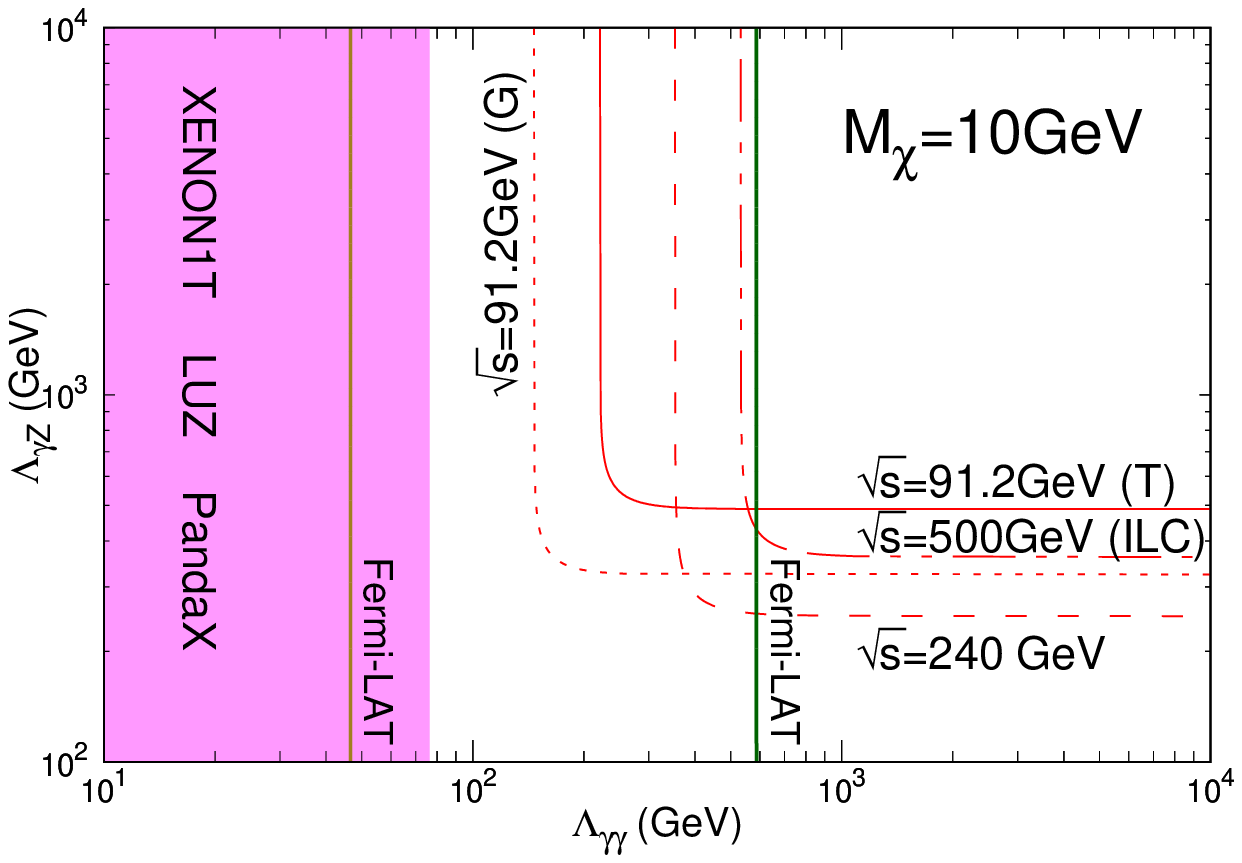}
     \hspace{0.01\textwidth}%
   \caption{limit on $\Lambda_{\gamma\gamma}-\Lambda_{\gamma Z}$ for direct, indirect and collider detection. In the left(right) panel, the DM is scalar (fermion) with show-case masses $m_{\phi, \chi}$ at 4 and 10 GeV. The cyan, magenta dash areas denote direct detection constraints for scalar DM and fermion DM with scalar type. The constraints from direction detection are SuperCDMS, CDEX, CDMSlite, XENON1T, LUX and PandaX. The purple, olive, dark green solid lines denote Fermi-LAT constraint from R3 region, where they denote scalar DM, fermion DM with scalar and pseudo-scalar types, respectively. The red (blue) solid/dotted/dashed line denotes fermion DM with scalar type (scalar DM) for CEPC $3\sigma$ sensitivities with unpolarized beams at the integrated luminosity of ${\rm 2.5ab^{-1}, 25fb^{-1}}$ for $\sqrt{s}=91.2$GeV and ${\rm 5ab^{-1}}$ for $\sqrt{s}=240$GeV. The labels T and G in the legend denote tera ($10^{11}$) and giga ($10^{9}$) Z bosons at $Z$-pole. The dotted-dashed line denotes for the (unpolarized) ILC 3$\sigma$ sensitivity with integrated luminosity of 500fb$^{-1}$ at $\sqrt{s}=500$GeV.}
  \label{fig:Lambda_gammaz}
\end{figure}

Fig.~\ref{fig:Lambda_gammaz} shows the CEPC, direct and indirect detection limits on the $\Lambda_{\gamma\gamma}-\Lambda_{\gamma Z}$ plane. In the left(right) panel, the DM is scalar (fermion) with masses $m_{\phi, \chi}$ at 4 and 10 GeV. The direct detection calculation only includes $\Lambda_{\gamma\gamma}$ contribution. For the indirect constraint, $\Lambda_{\gamma Z}$ contribution is small and does not cause visible shape-change in the plotted parameter range. The cyan, magenta dash areas denote direct detection constraints for scalar DM and fermion DM with scalar type. The constraints from direction detection are SuperCDMS~\cite{Agnese:2014aze}, CDEX~\cite{Jiang:2018pic}, CDMSlite~\cite{Agnese:2015nto}, XENON1T~\cite{Aprile:2017iyp}, LUX~\cite{Akerib:2016vxi} and PandaX~\cite{Cui:2017nnn}. The purple, olive, dark green solid lines denote Fermi-LAT constraint from R3 region~\cite{Ackermann:2015lka}, where they denote scalar DM, fermion DM with scalar and pseudo-scalar types, respectively. The red (blue) solid/dotted/dashed line denotes fermion DM with scalar type (scalar DM) for CEPC $3\sigma$ sensitivities with integrated luminosity of ${\rm 2.5ab^{-1}, 25fb^{-1}}$ at $\sqrt{s}=91.2$GeV and ${\rm 5ab^{-1}}$ at $\sqrt{s}=240$GeV, corresponding to prospective $10^{11}, 10^{9}$ $Z$ boson and $10^6$ Higgs runs. The labels T and G in the legend denote tera ($10^{11}$) and giga ($10^{9}$) Z bosons at $Z$-pole. Note the difference between pseudo-scalar and scalar types from collider constraints at low mass is very small, as illustrated in Fig.~\ref{fig:Lambda}, and thus we do not show the pseudo-scalar constraint lines in Fig.~\ref{fig:Lambda_gammaz}. For comparison we include for ILC's 3$\sigma$ mono-photon sensitivity (dotted-dashed line) with integrated luminosity of 500 fb$^{-1}$ at $\sqrt{s}=500$GeV~\cite{Asner:2013psa}.

While $\Lambda_{\gamma\gamma}$ can be more tightly constrained at the indirect-detection experiments, CEPC can offer good $\Lambda_{\gamma Z}$ sensitivity in the $Z$-pole runs. For a low $m_{\phi, \chi}$, giga-Z (tera-Z) run can probe $\Lambda_{\gamma Z}$ to 1030 (1970) GeV for scalar DM, to 360 (540) GeV for fermion DM at 3$\sigma$ sensitivity. This limit is higher than the LHC 8 TeV constraints~\cite{Crivellin:2015wva} and lower than the 13 TeV LHC monophoton results~\cite{bib:LHC_monophoton} for dimension -7 operator with the $\gamma\gamma\chi\chi$ interaction.

\section{Conclusion}
\label{sect:conclusion}

In this work, we consider dimension -6 (scalar) and -7 (fermion) effective DM diboson operators and their test via the monophoton search channel at the CEPC. With a focus on the $Z$-pole energy, the effective DM couplings to the $Z$ boson can be accessed at large luminosity giga-Z and tera-Z runs. A DM mass below $M_Z/2$ allows for the three-body $Z\rightarrow \bar{\chi}\chi\gamma$ monophoton final state, where the photon is energetic and it recoils against a large MET. The major SM background $e^+e^-\rightarrow \bar{\nu}\nu\gamma$ is relatively small and is under good control with a transverse photon momentum cut. We adopt optimized photon $P_T$ cuts 35 (45) GeV at the $Z$-pole (240) GeV runs for scalar DM and photon $P_T$ cuts 30 (35) GeV at the $Z$-pole (240) GeV runs for fermion DM and derive the 3$\sigma$ sensitivity for the effective diboson couplings $\Lambda_{\gamma Z}$ and $\Lambda_{\gamma \gamma}$.


Best $\Lambda_{\gamma Z}$ sensitivity occurs at $Z$-pole due to on-resonance production of the $Z$ boson, where $\Lambda_{\gamma Z}$ contribution dominates. Proposed 25 fb$^{-1}$ (2.5 ab$^{-1}$) $Z$-pole luminosity runs can probe $\Lambda_{\gamma Z}$ to 1030 (1970) GeV for scalar DM, to 360 (540) GeV for fermion DM at 3$\sigma$ sensitivity in the low DM mass limit. 240 GeV run loses sensitivity in $\Lambda_{\gamma Z}$ as the center of mass energy moves away from $Z$-pole and a better sensitivity in $\Lambda_{\gamma\gamma}$ is obtained instead, and at 5 ab$^{-1}$ luminosity $\Lambda_{\gamma\gamma}$ can be probed to 590 (360) GeV for scalar (fermion) DM. Sensitivity for variant DM mass and $\Lambda_{\gamma\gamma},\Lambda_{\gamma Z}$ combinations are given in Figs.~\ref{fig:Lambda} and ~\ref{fig:Lambda_gammaz}.

We compare the CEPC's sensitivities to current constraints from direct and indirect dark matter searches. Limits from the latest experiments are shown in Fig.~\ref{fig:Lambda_gammaz}. Non-collider searches can be very sensitive to $\Lambda_{\gamma\gamma}$ and give a higher than TeV $\Lambda_{\gamma\gamma}$ constraint in their optimal DM mass range. In comparison, the CEPC runs give better $\Lambda_{\gamma Z}$ sensitivity for DM masses accessible to the CEPC.

\bigskip
{\bf Acknowledgements}

\medskip
The authors thank Manqi Ruan, Xiao-jun Bi and Pengfei Yin for helpful discussions. Y.G. is supported under grant no.~Y7515560U1 by the Institute of High Energy Physics, Chinese Academy of Sciences. M.J. thanks the Institute of High Energy Physics, Chinese Academy of Sciences for support.

\onecolumngrid
\appendix


\renewcommand{\theequation}{A-\arabic{equation}}
\setcounter{equation}{0}

\section{Comparison of the nucleon mass $m_N$ and the reduced nucleon mass $\mu_N$ in direct detection}

The spin-independent DM-nucleon cross section is given in Ref.~\cite{Weiner:2012cb}, and a factor of 2 in the coefficients is corrected by Ref.~\cite{Frandsen:2012db},
\be
\sigma_{n} = \frac{8m^2_A\alpha^2Z^4Q^2_0F^2_{{\rm Ray}}(\bar{q})}{\pi^2A^4\Lambda_{\gamma\gamma}^6},
\label{eq:a1}
\ee
with the nucleus mass $m_{\rm A}=A\cdot m_n$, the nuclear coherence scale $Q_0\simeq 0.48(0.3+0.89A^{1/3})^{-1}$GeV, and the charge form factor $F_{{\rm Ray}}(\bar{q})$ at low momentum transfer is $F_{{\rm Ray}}(0)=1$.

The isotope-averaged A for Xenon (Germanium) is 131.3 (72.6). The reduced mass $\mu_{\rm A}=m_{\rm A}m_{\phi, \chi}/(m_{\rm A}+m_{\phi, \chi})$ can approximate to the smaller of $m_{\rm A}$ and $m_{\phi,\chi}$ if the two masses are very different. In the heavy DM limit, $\mu_A\approx m_{\rm A}$. In our case at the CEPC the relevant DM range is light in comparison to the nucleon mass, and we use the exact $\mu$ to calculate the scattering cross-section, as is given in Eq.~\ref{eq:dd7}. In Table~\ref{tab:mu_m}, we list the difference of the $\Lambda^{(\mu_A)}_{\gamma\gamma}$ and $\Lambda^{(m_A)}_{\gamma\gamma}$ by using exact $\mu$ and that from a $\mu\rightarrow m_{\rm A}$ approximation, and the latter would over-estimate the scattering rate at relatively low DM masses.

\begin{table}[h]
\begin{center}
\renewcommand{\arraystretch}{1.2}
\setlength\tabcolsep{0.5em}
\begin{tabular}{cccc|cccc}
\hline\hline
Ge& $m_{\chi}$ & $\Lambda^{(\mu_A)}_{\gamma\gamma}$ & $\Lambda^{(m_A)}_{\gamma\gamma}$ & 
Xe& $m_{\chi}$ & $\Lambda^{(\mu_A)}_{\gamma\gamma}$ & $\Lambda^{(m_A)}_{\gamma\gamma}$\\
\hline
                 & 5     &  13.8  &   34.5    &                  &  $10^1$  &   76.5   &  185     \\
                 & 10   &  33.3  &   67.2    &                  &  $10^2$  &   221    &   292    \\
                 & 20   &  47.7  &   79.4    &                  &  $10^3$  &   192    &   200    \\
                 & 30   &  52.5  &   79.2    &                  &  $10^4$  &   136    &   136    \\
 
\hline\hline
\end{tabular}
\end{center}
\caption{Comparison of $\Lambda_{\gamma\gamma}$ constraint by using $\mu_A$ or $m_A$. $\mu_A$ should be used for the low DM mass range that is favored by $Z$ pole searches.}
\label{tab:mu_m}
\end{table}

\end{document}